\documentclass[12pt]{article}
\usepackage{amsmath}
\usepackage{graphicx}
\usepackage{xcolor}
\usepackage{authblk} 

\usepackage[utf8]{inputenc}
\usepackage[numbers, compress]{natbib}
\usepackage{natbib}
\usepackage{slashed}
\usepackage{epsfig}

\title{Cherenkov radiation constraint on the photon mass from a two-form gauge potential}

\author[1]{Patricio Gaete}
\author[2]{Jos\'{e} Helay\"{e}l-Neto}
\affil[1]{Departamento de Física and Centro Científico-Tecnológico de Valparaíso (CCTVal), Universidad Técnica Federico Santa María, Valparaíso, Chile} 
\affil[2]{Centro Brasileiro de Pesquisas F\'{i}sicas (CBPF), Rio de Janeiro, RJ, Brasil}

\date{\today} 

\begin{document}

\maketitle
\begin{abstract}
We investigate how electromagnetic radiation in massive antisymmetric tensor theories can be used to constrain the photon mass. In this context, massive electrodynamics constitutes one realization of the $B \wedge F$ class of models in $(3+1)$ dimensions and is obtained by dualizing Stueckelberg-like massive gauge theories. Notably, this model is characterized by the mixing of a $U(1)$ gauge potential with an Abelian $3$-form field of the type found in the topological sector of QCD. We then examine how mass affects permittivity and susceptibility and, consequently, modifies the refractive index through the dispersion relation.
We subsequently investigate the electromagnetic radiation emitted by a moving charged particle as it interacts with a medium within the framework of massive electrodynamics. Our findings indicate that the medium through which the particle travels drives the radiation, in a manner analogous to the Cherenkov effect. Applying our results to data on Cherenkov radiation produced at Super-Kamiokande, we derive a photon mass bound of 1 eV ($ 10^ {- 36} $ kg), which is consistent with known estimates based on superconductors.
\end{abstract}

\section{Introduction}

Cherenkov radiation occurs when a charged particle travels through a medium at a speed exceeding the phase velocity of light in that medium, causing the emission of electromagnetic radiation. It was first observed experimentally by Cherenkov in 1934 \cite{Cherenkov} and subsequently explained theoretically by Frank and Tamm \cite{Tamm}. Since then, its physical implications have been explored from multiple perspectives, including applications in high-energy particle detection \cite{Ypsilantis,Abashian,Adam}, light sources \cite{Adamo,Liu,Ye}, metamaterials and photonic crystals \cite{Luo}, and chiral matter \cite{Urrutia}.

On the other hand, despite intensive efforts over more than three decades, a complete understanding of the QCD vacuum structure and the mechanism of color confinement remains elusive. String theory and the AdS/CFT duality \cite{Maldacena} provide one of the most promising technical frameworks for addressing nonperturbative QCD effects. Primarily motivated by interest in string-theoretic ideas, this approach has been extended beyond four dimensions to encompass theories of antisymmetric tensors of arbitrary rank in arbitrary spacetime dimensions, which appear as low-energy effective field theories of strings. This generalization has also yielded valuable insight into confinement mechanisms in a variety of settings.

In addition, antisymmetric tensor theories have long been viewed as natural extensions of Abelian gauge theories (our basic framework for gauge interactions). They share several key properties, most notably the presence of dualities, including the strong-weak coupling mappings familiar from electromagnetism. Antisymmetric tensors arise naturally in supersymmetric field theories and string theory, where they play a central role in realizing strong- and weak-coupling dualities among string theories. As we emphasized in our previous paper, confinement is a strong-coupling phenomenon, and there is no evidence that the strong- and weak-coupling phases are related by a smooth change of the coupling constant. In this work, we will use a duality that enables us to move between these two phases. In other words, in the strong-coupling phase, QCD is described by an Abelian three-index gauge potential rather than by the Yang-Mills field. Consequently, in different phases, the same physical system is characterized by different degrees of freedom. In fact, Ref. \cite{Gabadadze} provides a detailed analysis of the relationship between these two phases and explicitly derives the connection between the Abelian field $A_{\mu\nu\rho}$ and the gluon field.

In this work, we focus on the dualization of Stueckelberg-like massive gauge theories and $B \wedge F$ models, derived from a general $p$-form dualization framework for interacting theories in $d$ spacetime dimensions \cite{Ansoldi, Spallucci1,Spallucci2,Marinatto,Spallucci3}. In particular, in $(3+1)$ dimensions, we obtain the following $B \wedge F$ model:
\begin{equation}
{\cal L} =  - \frac{1}{4}F_{\mu \nu }^2\left( A \right) + \frac{1}{{12}}H_{\mu \nu \rho }^2\left( B \right) + \frac{M}{{24}}{\varepsilon ^{\mu \nu \rho \sigma }}{B_{\mu \nu }}{\partial _{[\rho }}{A_{\sigma ]}}.  \label{II-04}
\end{equation}
In the next section, we will show that this expression is equivalent to the Proca Lagrangian.

Our work is organized as follows. Section II reviews the key aspects of dualization using a simple Lagrangian and shows that it is equivalent to a massive Proca field. In this section, we also derive a nonlocal formulation of massive electrodynamics. We then compute the photonic dispersion relation to investigate how the mass term alters the vacuum's constitutive properties and determine the corresponding refractive indices, demonstrating that the resulting effective medium is dispersive. Based on these results, Section III examines the electromagnetic radiation from a moving charge interacting in this dispersive medium and evaluates the radiated energy using the standard Poynting vector approach. Using the foregoing information, we proceed to derive a bound on the photon mass. Finally, Section IV provides our concluding remarks.

In our conventions, the signature of the metric is ($+1,-1,-1,-1$).

\section{Model under consideration: dispersion relations and refractive indices}

We begin by briefly reviewing previous results on duality symmetry \cite{Aurilia}. Duality symmetry in gauge theories plays a central role because it provides equivalent descriptions of the same physical phenomena in different theoretical frameworks.
As a concrete setting, we focus on the dualization of Stueckelberg-like massive gauge theories and $B \wedge F$ models. This dualization follows from a general $ p$-dualization procedure for interacting theories in $d$ spacetime dimensions \cite{Spallucci1}. Specializing to $(3+1)$ dimensions, the construction yields the following $B \wedge F$ model:
\begin{equation}
{\cal L} =  - \frac{1}{4}F_{\mu \nu }^2\left( A \right) + \frac{1}{{12}}H_{\mu \nu \rho }^2\left( B \right) + \frac{M}{{24}}{\varepsilon ^{\mu \nu \rho \sigma }}{B_{\mu \nu }}{\partial _{[\rho }}{A_{\sigma ]}}.  \label{II-05}
\end{equation}

We next observe that the Lagrangian density in Eq. (\ref{II-05}) can be rewritten as follows:
\begin{equation}
{\cal L} =  - \frac{1}{4}F_{\mu \nu }^2 - \frac{1}{2}{\tilde H_\sigma }{\tilde H^\sigma } - \frac{m}{6}{\tilde H^\sigma }{A_\sigma }, \label{II-10}
\end{equation}
where we have introduced the dual field ${\tilde H^\mu } = {\raise0.5ex\hbox{$\scriptstyle 1$}\kern-0.1em/\kern-0.15em\lower0.25ex\hbox{$\scriptstyle 2$}}{\varepsilon ^{\mu \nu \lambda \rho }}{\partial _\nu }{B_{\lambda \rho }}$.
Eliminating the dual-field $H^{\sigma}$ requires special care, since it must satisfy the constraint ${\partial _\mu }{\tilde H^\mu } = 0$, which is the Bianchi identity. To implement this constraint, we introduce a Lagrange multiplier $\chi$. After some algebra, the effective Lagrangian density (\ref{II-10}) takes the form 
\begin{equation}
{\cal L} = - \frac{1}{4}F_{\mu \nu }^2 - \frac{1}{2}{\tilde H_\sigma }{\tilde H^\sigma } - \frac{M}{6}{\tilde H^\sigma }{A_\sigma } + \chi {\partial _\sigma }{\tilde H^\sigma }   .\label{II-15}
\end{equation}
Defining ${Z_\sigma } \equiv {A_\sigma } + \frac{6}{M}{\partial _\sigma }\chi$, with $  {Z_{\mu \nu }} = {F_{\mu \nu }}$, we obtain
\begin{equation}
{\cal L} =  - \frac{1}{4}Z_{\mu \nu }^2 - \frac{1}{2}{\tilde H_\sigma }{\tilde H^\sigma } - \frac{M}{6}{\tilde H^\sigma }{Z_\sigma }.\label{II-20} 
\end{equation}
If we further redefine the fields, as ${W_\sigma } \equiv {\tilde H_\sigma } + \frac{M}{6}{Z_\sigma }$, the Lagrangian density in Eq. (\ref{II-05}) can be rewritten as follows
\begin{equation}
{\cal L}=  - \frac{1}{4}Z_{\mu \nu }^2 + \frac{1}{2}{\mu ^2}Z_\mu ^2, \label{II-25}
\end{equation}
with ${\mu ^2} \equiv {\raise0.5ex\hbox{$\scriptstyle {{M^2}}$}
\kern-0.1em/\kern-0.15em
\lower0.25ex\hbox{$\scriptstyle {36}$}}$. It is immediately apparent that this expression closely resembles the Proca Lagrangian, which breaks gauge invariance. In the Appendix, we demonstrate that the sequence of field redefinitions leading to the massive photon leaves the two-form gauge field decoupled from the Z-field, identified as the (massive) photon. The Abelian B-field, which carries one physical degree of freedom on shell, provides the Z-field with a longitudinal component, analogous to the Higgs mechanism. 

Before proceeding further, we first demonstrate how gauge invariance is restored in Eq. (\ref{II-25}). To this end, we employ the Hamiltonian formalism for constrained systems as described in 
Ref. \cite{PGaete}. The canonical conjugate variables are given by $\Pi ^0=0$ and $\Pi ^i = - F^{0i}$. The canonical Hamiltonian is then,
\begin{equation}
{H_C} = \int {{d^3}x} \left[ { - {A_0}\left( {{\partial _i}{\Pi ^i} + \frac{{{M^2}}}{2}{A^0}} \right) - \frac{1}{2}{\Pi _i}{\Pi ^i} + \frac{1}{4}{F_{ij}}{F^{ij}} - \frac{{{M^2}}}{2}{A_i}{A^i}} \right], \label{II-30}
\end{equation}
where we have redefined ${{Z}}^{\mu}$ as ${{A}}^{\mu}$ and $M$ such that ${{M}}^{2} = {\mu}^{2}$.
   
Next, since the primary constraint $\Pi ^0=0$ must hold at all times, we obtain the following secondary constraint
\begin{equation}
\Gamma \left( x \right) \equiv \partial _i \Pi ^i  + \frac{M^2}{2} A^0  = 0. \label{II-35}
\end{equation}
From the discussion above, we see that the constraints are second-class, as expected for a theory with an explicit mass term that breaks gauge invariance. To convert this second-class system into a first-class one, we extend the original phase space by introducing a canonical pair of fields $\theta$ and $ \Pi _\theta $, following Ref. \cite{PGaete}. In this extended space, we can then define a new set of first-class constraints:
\begin{equation}
\Lambda _1  \equiv \Pi _0  + \frac{M^2}{2} \theta, \label{II-40}
\end{equation}
and
\begin{equation}
\Lambda _2  \equiv \Gamma  + \Pi _\theta. \label{II_45}
\end{equation}
In this way, the gauge symmetry of the theory is restored. After integrating out the $\theta$ field, the resulting effective Lagrangian is then given by
\begin{equation}
\mathcal{ L} = - \frac{1}{4}{F_{\mu \nu }}\left( {1 + \frac{{{M^2}}}{\Delta }} \right){F^{\mu \nu }}.
\label{II-50}
\end{equation}
Thus, equation (\ref{II-50}) yields a nonlocal formulation of  Proca theory.
In other words, in a vacuum, massive Proca fields are mathematically equivalent to Maxwell's theory formulated in a particular nonlocal (spatially dispersive) medium, known as Proca metamaterials (MTM).

To illustrate that the modified vacuum behaves like a material medium, we adopt the standard definitions ${\bf D} = \frac{{\partial {\cal L}}}{{\partial {\bf E}}}$, ${\bf H} =  - \frac{{\partial {\cal L}}}{{\partial {\bf B}}}$, along with the constitutive relations ${\bf D} = \varepsilon \,{\bf E}$ and ${\bf H} = {\mu ^{ - 1}}{\bf B}$. From these, we obtain the following expressions for the vacuum's effective permittivity and permeability:
\begin{equation}
\varepsilon = 1+ \frac{{{M}}^{2}}{\bigtriangleup}, \label{II-55}
\end{equation}
 and
 \begin{equation}
\mu = {\left({{1} + \frac{{{M}}^{2}}{\bigtriangleup}}\right)}^{{-}{1}}. \label{II-60}
 \end{equation}
From the preceding expressions, we note that the refractive index, defined as ${n^2} = \varepsilon\, \mu$, is equal to one. 

To clarify this point, we examine the dispersion relations (DRs) for an electromagnetic wave propagating through the modified electromagnetic vacuum. For this purpose, we decompose the electric and magnetic fields $\bf E$ and $\bf B$, respectively, into plane waves, as shown below:
\begin{equation}
{\bf E} = {{\bf E}_{0R}}\cos \left( {{\bf k} \cdot {\bf x} - \omega t} \right) - {{\bf E}_{0I}}sin\left( {{\bf k} \cdot {\bf x} - \omega t} \right), \label{II-65}
\end{equation}
and
\begin{equation}
{\bf B} = {{\bf B}_{0R}}\cos \left( {{\bf k} \cdot {\bf x} - \omega t} \right) - {{\bf B}_{0I}}sin\left( {{\bf k} \cdot {\bf x} - \omega t} \right). \label{II-70}
\end{equation}

It should also be noted that, in the absence of external sources, Maxwell's equations are given by:
\begin{eqnarray}
\left({{1}{+}\frac{{{M}}^{2}}{\bigtriangleup}}\right)\nabla \cdot {\bf E} = 
0, \nonumber\\
\nabla  \cdot {\bf B} = 0, \nonumber\\
\nabla  \times {\bf E} =  - \frac{{\partial \,{\bf B}}}{{\partial \, t}}, \nonumber\\
\left({{1}{+}\frac{{{M}}^{2}}{\bigtriangleup}}\right)\left[{\nabla \times{\bf B} -\frac{1}{{c^2}}\frac{\partial\,{\bf E}}{\partial\,{{t}}}}\right] = 0. 
\label{II-75}
\end{eqnarray}

Substituting equations (\ref{II-65}) and (\ref{II-70}) into equation (\ref{II-75}), yields
\begin{subequations}
\begin{eqnarray}
\left( {k^2} - {M^2} \right){\bf k} \cdot {{\bf E}_{0R}} = 0, \label{II-80} \\
\left( {k^2} - {M^2} \right){\bf k} \cdot {{\bf E}_{0I}} = 0, \label{II-85}
\end{eqnarray}
\end{subequations}
where ${k^2} = {k^\mu }{k_\mu } = \frac{{{\omega ^2}}}{{{c^2}}} - {{\bf k}^2}$.

We additionally have
\begin{subequations}
\begin{eqnarray}
{\bf k} \times {{\bf E}_{0R}} = \omega \, {{\bf B}_{0R}}, \label{II-90} \\
{\bf k} \times {{\bf E}_{0I}} = \omega \, {{\bf B}_{0I}}. \label{II-95}
\end{eqnarray}
\end{subequations}
Therefore, it follows that
\begin{equation}
{\bf B} = \left( {\frac{{\bf k}}{\omega } \times {{\bf E}_{0R}}} \right)\cos \left( {{\bf k} \cdot {\bf x} - \omega t} \right) - \left( {\frac{{\bf k}}{\omega } \times {{\bf E}_{0I}}} \right)\sin \left( {{\bf k} \cdot {\bf x} - \omega t} \right). \label{II-100}
\end{equation}

Finally, we have
\begin{subequations}
\begin{eqnarray}
\left( {{k^2} -{M^2} } \right){\bf k} \times {{\bf B}_{0R}} =  - \left( {{k^2} -{M^2 } } \right)\frac{\omega }{{{c^2}}}{{\bf E}_{0R}}, \label{BIR40}  \\
\left( {{k^2} - {M^2}} \right){\bf k} \times {{\bf B}_{0I}} =  - \left( {{k^2} -{M^2} } \right)\frac{\omega }{{{c^2}}}{{\bf E}_{0I}}. \label{II-105}
\end{eqnarray}
\end{subequations} 

Accordingly, the corresponding dispersion relations are expressed as follows:
\begin{equation}
{k^2}\left( {{k^2}-{M^2}} \right) = 0. \label{II-110}
 \end{equation}
However, the dispersion relation 
\begin{equation}
\frac{{{\omega ^2}}}{{{c^2}}} = {{\bf k}^2} + {M^2}, \label{II-115}
\end{equation}
implies that the phase velocity is
\begin{equation}
{v_{ph}} = \frac{\omega }{{|{\bf k} |}} > c, \label{II-120}
\end{equation}
and that the refractive index is
\begin{equation}
n = \frac{c}{{{v_{ph}}}} < 1. \label{II-125}
\end{equation}
Evidently, this is incompatible with the fact that the photon is massive, given that
\begin{equation}
\frac{{|{\bf v}|}}{c} = \frac{{|{\bf p}|\,c}}{E} < 1.  \label{II-130}
\end{equation}
Accordingly, we need to consider the group velocity
\begin{equation}
\frac{{{v_g}}}{c} = \frac{1}{c}\frac{{d\omega }}{{d|{\bf k}|}} = \frac{{|{\bf k}|\, c}}{\omega } < 1, \label{II-135}
\end{equation}
and the corresponding refractive index
\begin{equation}
n = \frac{c}{{{v_g}}} > 1, \label{II-140}
\end{equation}
so that
\begin{equation}
n{\left( \omega  \right)^2} = \frac{{{\omega ^2}}}{{|{\bf k}{|^2}{c^2}}} = \frac{{{\omega ^2}}}{{{\omega ^2} - {M^2}\,{c^2}}} > 1,
\end{equation}
with $\omega  > M\,c$.

Accordingly, the preceding electromagnetic vacuum behaves as a birefringent medium with refractive index $n$. There is, however, a further strand related to the existence of electromagnetic vacuum birefringence. Specifically, we consider the electromagnetic radiation emitted when a moving charged particle interacts with a medium within the framework of massive electrodynamics. 
In the following discussion, we will examine this phenomenon.
 
\section{Electromagnetic radiation}

We now examine the electromagnetic radiation emitted by a moving charged particle as it propagates through a medium governed by Proca-like electromagnetism.

Therefore, Maxwell's equations can be written as follows:
\begin{eqnarray}
\left({{1}{+}\frac{{{M}}^{2}}{\bigtriangleup}}\right)\nabla \cdot {\bf E} = 
\frac{{4\pi }}{\varepsilon }{\rho _{ext}}, \nonumber\\
\nabla  \cdot {\bf B} = 0, \nonumber\\
\nabla  \times {\bf E} =  - \frac{{\partial \,{\bf B}}}{{\partial \, t}}, \nonumber\\
\left({{1} + \frac{{{M}}^{2}}{\bigtriangleup}}\right)\left[{\nabla \times{\bf B} -\frac{\varepsilon\mu}{{c}}\frac{\partial\,{\bf E}}{\partial\,{{t}}}}\right] = \frac{{4}\pi\mu}{{c}}{\bf J}_{{ext}}. 
\label{III-05}
\end{eqnarray}

Here, ${\rho _{ext}}$ and ${{\bf J}_{ext}}$ denote the external charge and current densities, respectively, definied by: ${\rho _{ext}}\left( {t,{\bf x}} \right) = Q\,\delta \left( x \right)\delta \left( y \right)\delta \left( {z - {v}t} \right)$ and ${\bf J}_{ext}\left( {t,{\bf x}} \right) = Q\,{\bar u}\,\delta \left( x \right)\delta \left( y \right)\delta \left( {z - {v}t} \right){\hat {\bf e}_z}$. For simplicity, we choose the $z$-axis as the direction of motion of the charged particle, with ${v}$ denoting its velocity. Additionally, we define $\bigtriangleup = {\partial}_{\mu}{\partial}^{\mu}$. 

Note that the equations presented above can be rewritten in the following form:
\begin{subequations}
\begin{eqnarray}
\left({{\nabla}^{2} - \frac{1}{{{c}}^{\prime{2}}}\frac{{\partial}^{2}}{\partial{{t}}^{2}}}\right){{\bf E}} &=& \frac{{4}\pi}{\varepsilon}{\left({{1} + \frac{{{M}}^{2}}{\bigtriangleup}}\right)}^{-1}\nabla{\rho}_{{ext}} +\frac{{4}\pi\mu}{{{c}}^{2}}{\left({{1} + \frac{{{M}}^{2}}{\bigtriangleup}}\right)}^{-1}\frac{\partial{{\bf J}}_{{ext}}}{\partial{{t}}}, \nonumber\\ 
\label{III-10a}\\
&&\left({{\nabla}^{2} - \frac{1}{{{c}}^{\prime{2}}}\frac{{\partial}^{2}}{\partial{{t}}^{2}}}\right){{\bf B}} = - \frac{{4}\pi\mu}{{c}}{\left({{1} + \frac{{{M}}^{2}}{\bigtriangleup}}\right)}^{-1}\nabla\times{{\bf J}}_{{ext}}. \nonumber\\
\label{III-10b}
\end{eqnarray}
\end{subequations}
Here, we have simplified our notation by setting $\frac{1}{c'^2} \equiv \frac{\varepsilon \mu}{c^2}$.

As discussed in \cite{PGaete2}, we now perform a Fourier transform to momentum space:
\begin{equation}
F(t,{\bf x}) = \int {\frac{{d \omega {d^3}{\bf k}}}{{{{\left( {2\pi } \right)}^4}}}} {e^{ - i\,\omega t + i\,{\bf k} \cdot {\bf x}}}F\left( {\omega,{\bf k}} \right), \label{III-15}
\end{equation}
where $F$ represents the electric and magnetic fields. 

Once this step is completed, we obtain the following expressions for the electric and magnetic fields:
\begin{equation}
{\bf B}\left( {\omega ,{\bf k}} \right) = i\frac{{4\pi \mu }}{c}\frac{1}{{{\cal P}\,{\cal O}}}\left( {{\bf k} \times {{\bf J}_{ext}}} \right),  \label{III-20}
\end{equation}
and
\begin{equation}
{\bf E}\left( {\omega ,{\bf k}} \right) =  - i\frac{{4\pi }}{\varepsilon}\frac{1} {{{\cal P}\,{\cal O}}}\,{\bf k}\,{\rho _{ext}} + i\frac{{4\pi \mu }}{{{c^2}}}\frac{1}{{{\cal P}\,{\cal O}}}\,\omega\, {\bf J}_{ext},  \label{III-25}
\end{equation}
here $\cal P$ and $\cal O$ are expressed as  
\begin{equation}
{\cal P} = {{\bf k}^2} - \frac{{{\omega ^2}}}{{{c^{\prime 2}}}},  \label{III-30}
\end{equation}
and
\begin{equation}
{\cal O} = \left({{1} + \frac{{{M}}^{2}}{\bigtriangleup}}\right) =  1 +\frac{{{M}}^{2}}{\left({{\bf k}^{2} - \raise0.7ex\hbox{${{\omega}^{2}}$}\!\!\left/{}\right.\!\!\lower0.7ex\hbox{${{{c}}^{2}}$}}\right)}. \label{III-35}
\end{equation}

Similarly, in Fourier space, the external charge and current densities take the form: $\rho_{ext} \left( {\omega,{\bf k}} \right) = 2\pi Q\delta \left( {\omega - {k_z}v} \right)$ and ${{\bf J}_{ext}}\left( {\omega,{\bf k}} \right) = 2\pi Qv\delta \left( {\omega - {k_z}v} \right){\hat {\bf e}_z}$.
 
Following the approach in \cite{PGaete2}, we compute ${\bf b}\left( {w,{\bf x}} \right)$ and ${\bf e}\left( {w,{\bf x}} \right)$. In particular, the expression for ${\bf b}\left( {\omega,{\bf x}} \right)$ is 
\begin{equation} 
{\bf B}\left( {\omega,{\bf x}} \right) = \int {\frac{{{d^3}{\bf k}}}{{{{\left( {2\pi } \right)}^3}}}} \; {e^{i{\bf k} \cdot {\bf x}}}\;{\bf B}\left( {\omega,{\bf k}} \right). \label{III-40}
\end{equation}

Since the charged particle moves along the $z$-axis, it is natural to use cylindrical coordinates, which capture the axial symmetry of the problem. In these coordinates, the magnetic field given by equation (\ref{III-40}) takes the form:
\begin{eqnarray}
{\bf B}\left( {\omega ,{\bf x}} \right) &=&  \frac{{iQ}}{{\pi c}}{e^{i{{\omega z} \mathord{\left/
 {\vphantom {{\omega z} v}} \right.
 \kern-\nulldelimiterspace} v}}}\int_0^\infty  {d{k_T}}\, {k_T}\int_0^{2\pi } {d\alpha } \frac{{{e^{i{k_T}{x_T}\cos \alpha }}}}{{{{\left. {\left( {{{\bf k}^2} - \frac{{{n^2}{\omega ^2}}}{{{c^2}}}} \right)} \right|}_{{k_z} = {\omega  \mathord{\left/
 {\vphantom {\omega  v}} \right.
 \kern-\nulldelimiterspace} v}}}}} \nonumber\\
 &\times&\left[ {{k_T}sen\alpha\, \pmb {\hat \rho}  - {k_T}\cos \alpha\, \pmb {\hat \phi} } \right] \nonumber\\
&-&\frac{{iQ}}{{\pi c}}\frac{2{{M}}^{2}{{e^{i{{\omega z} \mathord{\left/{\vphantom {{\omega z} v}} \right.\kern-\nulldelimiterspace} v}}}}}{\left({{1}{-}{{n}}^{2}}\right){{\omega}^{2}}\slash{{{c}}^{2}}}
\int_0^\infty  {d{k_T}}\, {k_T}\int_0^{2\pi } {d\alpha } \frac{{{e^{i{k_T}{x_T}\cos \alpha }}}}{{{{\left. {\left( {{{\bf k}^2} - \frac{{{\omega ^2}}}{{{c^2}}}} \right)} \right|}_{{k_z} = {\omega  \mathord{\left/
 {\vphantom {\omega  v}} \right.
 \kern-\nulldelimiterspace} v}}}}} \nonumber\\
&\times& \left[ {{k_T}sen\alpha\, \pmb {\hat \rho}  - {k_T}\cos \alpha\, \pmb {\hat \phi} } \right] \nonumber\\
 &+&\frac{{iQ}}{{\pi c}}\frac{2{{M}}^{2}{{e^{i{{\omega z} \mathord{\left/{\vphantom {{\omega z} v}} \right.\kern-\nulldelimiterspace} v}}}}}{\left({{1}{-}{{n}}^{2}}\right){{\omega}^{2}}\slash{{{c}}^{2}}}
\int_0^\infty  {d{k_T}} \,{k_T}\int_0^{2\pi } {d\alpha } \frac{{{e^{i{k_T}{x_T}\cos \alpha }}}}{{{{\left. {\left( {{{\bf k}^2} - \frac{{ {n^2} {\omega ^2}}}{{{c^2}}}} \right)} \right|}_{{k_z} = {\omega  \mathord{\left/
 {\vphantom {\omega  v}} \right.
 \kern-\nulldelimiterspace} v}}}}} \nonumber\\
&\times& \left[ {{k_T}sen\alpha\, \pmb {\hat \rho}  - {k_T}\cos \alpha\, \pmb {\hat \phi} } \right]. 
 \label{III-45}
\end{eqnarray}
Here $\pmb{\hat \rho}$ and $\pmb{\hat \phi}$ denote the unit vector normal and tangential to the cylindrical surface, respectively. Additionally, the subscript $T$ in $k_{T}$ refers to the component perpendicular to the $z$ axis. Furthermore, keeping terms up to order $M^{2}$, we use the relation: \\
$\frac{\mu}{{\cal P \cal O}}{=}\frac{1}{\left({{\bf k}^{2}{-}\frac{{{n}}^{2}{\omega}^{2}}{{{c}}^{2}}}\right)}{-}\frac{2{{M}}^{2}}{\left({{1}{-}{{n}}^{2}}\right){{\omega}^{2}}\slash{{{c}}^{2}}}\left\{{\frac{1}{\left({{{\bf k}}^{2} - \frac{{\omega}^{2}}{{{c}}^{2}}}\right)}{-}\frac{1}{\left({{{\bf k}}^{2}{-}\frac{{{n}}^{2}{\omega}^{2}}{{{c}}^{2}}}\right)}}\right\}$.

Besides, note that $\int_0^{2\pi } {d\theta } {e^{ix\cos \theta }}\sin \theta  = 0$, and $\int_0^{2\pi } {d\theta } {e^{ix\cos \theta }}\cos \theta  = 2\pi i{J_1}\left( x \right)$,  where ${{J_1}\left( x \right)}$ denotes the  Bessel function of the first kind. Using these identities, equation (\ref{III-45}) simplifies to:
\begin{eqnarray}
{{\bf B}}\left({\omega{,}{{\bf x}}}\right) &=& \frac{2{Q}}{{c}}{{e}}^{{i}{\omega{{z}}}\slash{{v}}}
\left({{1} + \frac{2{{M}}^{2}}{\left({{1} - {{n}}^{2}}\right){{\omega}^{2}}\slash{{{c}}^{2}}}}\right)
\int_0^\infty{{{dk}}_{{T}}\,{{k}}_{{T}}^{2}}\,\frac{{{J}}_{1}\left({{{k}}_{{T}}{{x}}_{{T}}}\right)}{\left({{{k}}_{{T}}^{2}{-}{\alpha}^{2}}\right)} \pmb{\hat \phi} \nonumber\\
&-& \frac{2{Q}}{{c}}{{e}}^{{i}{\omega{{z}}}\slash{{v}}}
\frac{2{{M}}^{2}}{\left({{1}{-}{{n}}^{2}}\right){{\omega}^{2}}\slash{{{c}}^{2}}}
\int_0^\infty
{{{dk}}_{{T}}\,{{k}}_{{T}}^{2}}\,\frac{{{J}}_{1}\left({{{k}}_{{T}}{{x}}_{{T}}}\right)}{\left({{{k}}_{{T}}^{2}{-}{\beta}^{2}}\right)} \pmb{\hat \phi}, \label{III-50}
\end{eqnarray}
where ${\alpha ^2} = \frac{{{\omega ^2}{n^2}}}{{{c^2}}}\left( {1 - \frac{{{c^2}}}{{{n^2}{v^2}}}} \right)$ and ${\beta ^2} = \frac{{{\omega ^2}}}{{{c^2}}}\left( {1 - \frac{{{c^2}}}{{{v^2}}}} \right)$.

Next, integrating over $k_T$ and carrying out some additional manipulations, we find that equation (\ref{III-50}) is as follows
\begin{eqnarray}
{{\bf B}}\left({\omega{,}{{\bf x}}}\right)&=&\frac{{{i}}\pi{{Q}}}{{c}}{{e}}^{{i}{\omega{{z}}}\slash{{v}}}\left({{1}{-}\frac{2{{M}}^{2}}{\left({{{n}}^{2}{-}{1}}\right){{\omega}^{2}}\slash{{{c}}^{2}}}}\right)\alpha\,{{H}}_{1}^{\left({1}\right)}\left({\alpha\rho}\right)\pmb{\hat \phi} \nonumber\\
&+&{{i}}\pi{{Q\,c}}\,{{e}}^{{i}{\omega{{z}}}\slash{{v}}}\frac{2{{M}}^{2}}{\left({{{n}}^{2}{-}{1}}\right){\omega}^{2}}\,\beta\,{{H}}_{1}^{\left({1}\right)}\left({\beta\rho}\right) \pmb{\hat \phi}. \label{III-55}
\end{eqnarray}
In this context, when working in cylindrical coordinates, we set ${x_T} = \rho$. Additionally, ${H}_{1}^{(1)}(x)$ denotes a Hankel function of the first kind.

Next, we determine the electric field. To this end, we first note that, to order $M^2$, we can express 
$\frac{1}{{\varepsilon {\cal P}{\cal O}}} = \frac{1}{\left({{\bf k}^{2}{-}\frac{{{n}}^{2}{\omega}^{2}}{{{c}}^{2}}}\right)}{-}\frac{2{{M}}^{2}}{\left({{1}{-}{{n}}^{2}}\right){{\omega}^{2}}\slash{{{c}}^{2}}}\left\{{\frac{1}{\left({{{\bf k}}^{2}{-}\frac{{\omega}^{2}}{{{c}}^{2}}}\right)}{-}\frac{1}{\left({{{\bf k}}^{2}{-}\frac{{{n}}^{2}{\omega}^{2}}{{{c}}^{2}}}\right)}}\right\}$. Applying the same procedure as in the previous step, we obtain the following expression for the electric field 
\begin{eqnarray}
{\bf E}\left( {\omega ,{\bf x}} \right) &=& \frac{{2Q}}{v}{e^{i\omega {z \mathord{\left/
 {\vphantom {z v}} \right.
 \kern-\nulldelimiterspace} v}}}\left( {1 + \frac{{2{M^2}}}{{\left( {1 - {n^2}} \right){{{\omega ^2}} \mathord{\left/
 {\vphantom {{{\omega ^2}} {{c^2}}}} \right.
 \kern-\nulldelimiterspace} {{c^2}}}}}} \right)\int_0^\infty  {d{k_T}} \,k_T^2\frac{{{J_1}\left( {{k_T}{x_T}} \right)}}{{\left( {k_T^2 - {\alpha ^2}} \right)}} \pmb  {\hat \rho} \nonumber\\
 &-& \frac{{2Q}}{v}{e^{i\omega {z \mathord{\left/
 {\vphantom {z v}} \right.
 \kern-\nulldelimiterspace} v}}}\frac{{2{M^2}}}{{\left( {1 - {n^2}} \right){{{\omega ^2}} \mathord{\left/
 {\vphantom {{{\omega ^2}} {{c^2}}}} \right.
 \kern-\nulldelimiterspace} {{c^2}}}}}\int_0^\infty  {d{k_T}} \,k_T^2\frac{{{J_1}\left( {{k_T}{x_T}} \right)}}{{\left( {k_T^2 - {\beta ^2}} \right)}} \pmb  {\hat \rho} \nonumber\\
 &+& i2Q{e^{i\omega {z \mathord{\left/
 {\vphantom {z v}} \right.
 \kern-\nulldelimiterspace} v}}}\left( {1 + \frac{{2{M^2}}}{{\left( {1 - {n^2}} \right){{{\omega ^2}} \mathord{\left/
 {\vphantom {{{\omega ^2}} {{c^2}}}} \right.
 \kern-\nulldelimiterspace} {{c^2}}}}}} \right)\omega \left( {\frac{1}{{{c^2}}} - \frac{1}{{{v^2}}}} \right) \nonumber\\
 &\times&\int_0^\infty  {d{k_T}} \,{k_T}\frac{{{J_0}\left( {{k_T}{x_T}} \right)}}{{\left( {k_T^2 - {\alpha ^2}} \right)}}    \pmb{{{\hat e}_z}} \nonumber\\
& -& i2Q{e^{i\omega {z \mathord{\left/
 {\vphantom {z v}} \right.
 \kern-\nulldelimiterspace} v}}}\frac{{2{M^2}}}{{\left( {1 - {n^2}} \right){{{\omega ^2}} \mathord{\left/
 {\vphantom {{{\omega ^2}} {{c^2}}}} \right.
 \kern-\nulldelimiterspace} {{c^2}}}}}\omega \left( {\frac{1}{{{c^2}}} - \frac{1}{{{v^2}}}} \right)\int_0^\infty  {d{k_T}} \,{k_T}\frac{{{J_0}\left( {{k_T}{x_T}} \right)}}{{\left( {k_T^2 - {\beta ^2}} \right)}}
\pmb{{{\hat e}_z}}. \nonumber\\  
\label{III-60}
\end{eqnarray}
Here, $\boldsymbol{\hat{e}_z}$ is the unit vector along the $z$ axis. Using the standard identity $\int_0^{2\pi} d\theta\ e^{ix\cos\theta} = 2\pi J_0(x)$, where $J_0(x)$ is the Bessel function of the first kind, and  then integrating over $k_T$, we can rewrite equation (\ref{III-60}) as follows:
\begin{eqnarray}
{\bf E}\left( {\omega ,{\bf x}} \right) &=& i\frac{{Q\pi }}{v}{e^{i\omega {z \mathord{\left/
 {\vphantom {z v}} \right.
 \kern-\nulldelimiterspace} v}}}\left( {\frac{{{\alpha ^2} - {\beta ^2} - 2{M^2}}}{{\left( {{n^2} - 1} \right){{{\omega ^2}} \mathord{\left/
 {\vphantom {{{\omega ^2}} {{c^2}}}} \right.
 \kern-\nulldelimiterspace} {{c^2}}}}}} \right)\alpha H_1^{\left( 1 \right)}\left( {\alpha \rho } \right)\pmb  {\hat \rho}\nonumber\\
 &+& i\frac{{Q\pi }}{v}{e^{i\omega {z \mathord{\left/
 {\vphantom {z v}} \right.
 \kern-\nulldelimiterspace} v}}}\frac{{2{M^2}}}{{\left( {{n^2} - 1} \right){{{\omega ^2}} \mathord{\left/
 {\vphantom {{{\omega ^2}} {{c^2}}}} \right.
 \kern-\nulldelimiterspace} {{c^2}}}}}\beta H_1^{\left( 1 \right)}\left( {\beta \rho } \right) \pmb  {\hat \rho} \nonumber\\
 &-& \pi Q{e^{i\omega {z \mathord{\left/
 {\vphantom {z v}} \right.
 \kern-\nulldelimiterspace} v}}}\left( {\frac{{{\alpha ^2} - {\beta ^2} - 2{M^2}}}{{\left( {{n^2} - 1} \right){{{\omega ^2}} \mathord{\left/
 {\vphantom {{{\omega ^2}} {{c^2}}}} \right.
 \kern-\nulldelimiterspace} {{c^2}}}}}} \right)\frac{{{\beta ^2}}}{\omega }H_0^{\left( 1 \right)}\left( {\alpha \rho } \right)  \pmb{{{\hat e}_z}} \nonumber\\
 &-& \pi Q{e^{i\omega {z \mathord{\left/
 {\vphantom {z v}} \right.
 \kern-\nulldelimiterspace} v}}}\frac{{2{M^2}}}{{\left( {{n^2} - 1} \right){{{\omega ^2}} \mathord{\left/
 {\vphantom {{{\omega ^2}} {{c^2}}}} \right.
 \kern-\nulldelimiterspace} {{c^2}}}}}\frac{{{\beta ^2}}}{\omega }H_0^{\left( 1 \right)}\left( {\beta \rho } \right) \pmb{{{\hat e}_z}}.   \label{III-65}
\end{eqnarray}

With this information, we can now determine the radiated energy in the case under consideration. As noted earlier, our approach is to estimate radiated energy by computing the Poynting vector.
We begin by recalling that the power density carried by the radiation fields through the surface $S$, enclosing the volume $V$, is given by the real part of the time-averaged Poynting vector.
\begin{equation}
{\bf S} = \frac{c}{{2\pi }}{\mathop{\rm Re}\nolimits} \left( {{\bf E} \times {{\bf B}^ * }} \right). \label{III-70}
\end{equation}
Specifically, we will calculate the power radiated through the surface, which can be expressed as
 \begin{equation}
{\cal E} = \int_{ - \infty }^\infty \! {dt} \int\limits_S {d{\bf a} \cdot {\bf S}}.  \label{III-75}
\end{equation}
With this in mind, we consider a cylinder of length $h$ and radius $\rho$. The total energy radiated through its surface can therefore be written as follows:
\begin{equation}
{\mathcal{E}} = \frac{c}{{2\pi }}\left( {2\pi \rho h} \right)\int_0^\infty  {d\omega }\, {S_\rho }\,\Theta \left( {v - \frac{c}{n}} \right). \label{III-80}
\end{equation}
As discussed previously, it is also important to note that we include the step function because Cherenkov radiation is emitted only when the particle's velocity exceeds the speed of light in the medium (${v} > \frac{c}{n}$), with $n > 1$.

In the radiation zone, we can use equations (\ref{III-55}) and (\ref{III-65}) to express the power radiated per unit length as follows:
\begin{eqnarray}
W \equiv \frac{\cal E}{h} &=& \frac{{2\,\pi\, {Q^2}}}{{{c^2}}}\left( {1 - \frac{{{c^2}}}{{{v^2}}}} \right)\int_0^\infty  {d\omega }\, \frac{{{n^2}\left( \omega  \right)}}{{\left( {{n^2}\left( \omega  \right) - 1} \right)}}\,\omega \left( {1 - \frac{{{c^2}}}{{{n^2}\left( \omega  \right){v^2}}}} \right) \nonumber\\
&\times&\left[ {1 - \frac{{2\,{c^2}{M^2}}}{{{\omega ^2}\left( {{n^2}\left( \omega  \right) - 1} \right)}}} \right]. \label{III-85}
\end{eqnarray}
It is worth noting that expression (\ref{III-85}) is one of the several terms that arise in the calculation of the $\rho$ component of the Poynting vector.
We also note that, to obtain the final expression (\ref{III-85}), we used the fact that in the radiation zone ($\alpha \rho  \gg 1$), the Hankel functions take the form
\begin{equation}
H_0^{\left( 1 \right)}\left( {\alpha \rho } \right) \approx \sqrt {\frac{2}{{\pi \alpha \rho }}} \,{e^{i\alpha \rho }}{e^{ - i{\pi  \mathord{\left/
 {\vphantom {\pi  4}} \right.
 \kern-\nulldelimiterspace} 4}}}, \label{III-90}
\end{equation}
\begin{equation}
H_1^{\left( 1 \right)}\left( {\alpha \rho } \right) \approx \sqrt {\frac{2}{{\pi \alpha \rho }}} \,{e^{i\alpha \rho }}{e^{ - i3{\pi  \mathord{\left/
 {\vphantom {\pi  4}} \right.
 \kern-\nulldelimiterspace} 4}}}. \label{III-95}
\end{equation}

Finally, we observe that, by substituting ${\omega _M} = c\, M$, equation (\ref{III-85}) can be rewritten as
\begin{equation}
W = \frac{{6\pi {Q^2}}}{{{c^2}}}\left( {1 - \frac{{{c^2}}}{{{v^2}}}} \right)\int_0^\infty  {d\omega \omega \left( {1 - \frac{{{c^2}}}{{{n^2}\left( \omega  \right){v^2}}}} \right)},  \label{III-100}
\end{equation}
where, based on Eq. (\ref{III-85}), we again selected a term analogous to Cherenkov radiation.

It should be noted that the preceding expression for the energy lost per unit length is expressed as an integral over frequency. However, the corresponding range of integration is very narrow, extending between two frequencies, $\omega_{1}$
and $\omega_{2} = \omega_{1} + \Delta \,{\omega}$.
Using $1 - \frac{1}{{{\beta ^2}}} =  - \frac{1}{{{\gamma ^2} - 1}}$ and $\frac{1}{{{\beta ^2}}} = \frac{{{\gamma ^2}}}{{{\gamma ^2} - 1}}$, equation  (\ref{III-100}) can be rewritten as
\begin{equation}
W =  - 6\pi \frac{{{Q^2}}}{{{c^2}}}\frac{1}{{{\gamma ^2} - 1}}\int_{{\omega _1}}^{{\omega _2}} {d\omega } \omega \left( {1 - \frac{{{\gamma ^2}}}{{{\gamma ^2} - 1}}\left( {1 - \frac{{{c^2}{M^2}}}{{{\omega ^2}}}} \right)} \right). \label{III-105}
\end{equation}
We also note, in passing, that $M$ is the inverse of the photon's Compton wavelength,
\begin{equation}
M = \mathchar'26\mkern-10mu\lambda _\gamma ^{ - 1} = \frac{{{m_\gamma }c}}{\hbar }, \label{III-110}
\end{equation}
which we expect to be very small, given the photon’s very low mass.

In proceeding with our estimation of $M$, $W$ may be expressed as follows:
\begin{equation}
W =  - 6\pi \frac{{{Q^2}}}{{{c^2}}}\frac{1}{{{{\left( {{\gamma ^2} - 1} \right)}^2}}}\int_{{\omega _1}}^{{\omega _1} + \Delta \omega } {d\omega } \left( { - \omega  + {\gamma ^2}\frac{{{c^2}{M^2}}}{\omega }} \right). \label{III-115}
\end{equation}
We therefore obtain the following expression for $W$:
\begin{equation}
W =  - 6\pi \frac{{{Q^2}}}{{{c^2}}}\frac{1}{{{{\left( {{\gamma ^2} - 1} \right)}^2}}}\left[ { - \left( {{\omega _1} + \frac{{\Delta \omega }}{2}} \right)\Delta \omega  + {\gamma ^2}{c^2}{M^2}\ln \frac{{{\omega _1} + \Delta \omega }}{{{\omega _1}}}} \right]. \label{III-120}
\end{equation}
Next, to illustrate the calculation numerically, we consider frequencies in the visible spectrum: ${\omega _1} = 0.5 \times {10^{15}}Hz$ (wavelength $\sim 600$ nm) and ${\omega _2} = {\omega _1} + \Delta \omega  = 1.0 \times {10^{15}}Hz$ (wavelength $\sim 300$ nm). In this case, $\ln \frac{{{\omega _1} + \Delta \omega }}{{{\omega _1}}}$ is of ${\cal O}\left( 1 \right)$, and $M^{2}$ may be expressed as follows:
\begin{equation}
{M^2} \sim \frac{1}{{{c^2}{\gamma ^2}}}\left[ {\frac{{{c^2}}}{{6\pi {Q^2}}}{{\left( {{\gamma ^2} - 1} \right)}^2}|W| - \frac{3}{2}\omega _1^2} \right]. \label{III-125}
\end{equation}
To estimate $M$, and, consequently, the photon mass, $m_{\gamma}$, we select data from neutrino detectors, such as Super-Kamiokande and IceCube. In this scenario, electrons propagate through water ($n\sim1.33$) with a kinetic energy of $1MeV$. At this energy, $\beta  \sim 0.94$, $\gamma  \sim 2.96$, and the energy loss is $W \sim 0.15\,{{keV} \mathord{\left/{\vphantom {{keV} {cm}}} \right.\kern-\nulldelimiterspace} {cm}}$, with $Q = e =  - \,4.8 \times {10^{ - 10}}\,stat\,C$.

With the previous values,  $\beta n > 1$, so we ensure that radiation is emitted. Taking these data into account, we estimate
\begin{equation}
{M^2} \sim {10^8}\,c{m^{ - 2}}, \label{III-130}
\end{equation}
and 
\begin{equation}
{\lambda _\gamma } \sim 1\,\mu\, m.  \label{III-135}
\end{equation}
Making use of ${m_\gamma } = \frac{\hbar }{c}M$, we get
\begin{equation}
{m_\gamma } \sim {10^{ - 33}}g\,\,\left( { \sim 1\,eV} \right). \label{III-135}
\end{equation}

It is also noteworthy that the paper Ref. \cite{Gruzinov} recently examined a photon mass generated through a Higgs mechanism involving a light, millicharged Higgs boson and estimated the photon mass to be $10^{-5}eV$. We will revisit this topic in our Concluding Remarks.

\section{Concluding Remarks}

In the present contribution, we adopt the Cherenkov radiation scenario to re-assess the issue of setting bounds on the photon mass. The mass-generation mechanism is triggered by the mixing of two Abelian fields: a vector field and a two-form gauge potential. No Higgs field is present to spontaneously break the $U(1)$ symmetry associated with the vector potential. Instead, the two-form field, which is on-shell equivalent to a scalar, is responsible for transferring its degree of freedom to the longitudinal component of the $U(1)$ gauge potential. As discussed in the Appendix, electrons remain decoupled from the two-form potential after the field redefinitions we perform. We show how the photon mass contributes to the energy dissipated per unit length through the Cherenkov effect. This approach yields an estimate of $1 eV$ for the photon mass. 

   In ref. \cite{Gruzinov} presents and discusses an interesting model based on treating the Universe as a superconductor. The authors estimate the photon mass to be $10^ {-5} eV$, arising from symmetry breaking induced by a light, millicharged Higgs scalar. In our case, the two-form potential plays the role of the millicharged Higgs in Ref. \cite{Gruzinov}. It is well known that, in scenarios involving massive photons, the mass is not universal; rather, its value depends on the particular physical system under consideration. In conventional superconductors, high-$T_c$ superconductors, specific models involving Lorentz-symmetry violation —such as the Carroll-Field-Jackiw model— deviations from the Ampère-Maxwell model, and other settings, the photon mass takes different values. In the specific case of the Cherenkov radiation detected in Super-Kamiokande, as discussed in Section $3$, the data used to obtain our estimate are of the order of $1 eV$.
   
   Finally, we have in mind to apply our study of Cherenkov radiation in the presence of massive photons to investigate the issue of energy loss in flares emitted by Blazars. These compact objects provide an excellent laboratory for studying several aspects of high-energy (TeV and PeV) astrophysics, including emission processes and particle acceleration. These flares may occur across different wavelength and energy bands, offering a promising avenue for research into Cherenkov radiation involving massive photons.

\section{Appendix}

As demonstrated in Section $2$, only the gauge sector was considered. In this Appendix, however, we minimally couple the electron field to the $U(1)$ gauge potential, $A^{\mu}$. The electron is neutral under the standard $U(1)$ symmetry associated with the rank-$2$ gauge potential. We aim to demonstrate that the field redefinitions carried out in Section $2$ leave the electron minimally coupled exclusively to the $Z^{\mu\nu}$ field, which is interpreted as the massive photon in accordance with the Lagrangian density in Eq. (\ref{II-25}).

Accordingly, returning to Eq. (\ref{II-05}), we define the electron field, $\Psi$, as follows:
\begin{eqnarray}
{\cal L} &=&  - \frac{1}{4}F_{\mu \nu }^2\left( A \right) + \frac{1}{{12}}H_{\mu \nu \rho }^2\left( B \right) + \frac{M}{{24}}{\varepsilon ^{\mu \nu \rho \sigma }}{B_{\mu \nu }}{\partial _{[\rho }}{A_{\sigma ]}} \nonumber\\
 &+& \bar \Psi i{\gamma ^\mu }\left( {{\partial _\mu } + iQ{A_\mu }} \right)\Psi  - m\bar \Psi \Psi.
\label{App05} 
\end{eqnarray}
Next, we introduce the Lagrange multiplier, $\chi$, and redefine the field as follows:
\begin{equation}
{Z_\mu } \equiv {A_\mu } + \frac{6}{M}{\partial _\mu }\chi. \label{App10}  
\end{equation}
In this manner, the fermion couplings were found to be 
\begin{equation}
- Q\bar \Psi {\gamma ^\mu }\Psi {Z_\mu } + \frac{6}{M}Q\bar \Psi {\gamma ^\mu }\Psi {\partial _\mu }\chi. \label{App15}  
\end{equation}
However, after partial integration and the application of current conservation, the coupling to the Lagrange multiplier vanishes. Consequently, the fermion couples to the $Z^{\mu}$ field, which we interpret as the massive photon. Therefore, Eq. (\ref{II-25}) should be written as follows:
\begin{equation}
{\cal L} =  - \frac{1}{4}Z_{\mu \nu }^2 + \frac{1}{2}{\mu ^2}Z_\mu ^2 + \bar \Psi i{\gamma ^\mu }\left( {{\partial _\mu } + iQ{Z_\mu }} \right)\Psi  - m\bar \Psi \Psi. \label{App20} 
\end{equation}
In the present paper, we use this Lagrangian model to study Cherenkov radiation.

\section*{Acknowledgments}

The authors wish to express their gratitude to their colleague, Denis Cocuroci, for the discussions and for recommending relevant references.


\begin{thebibliography}{99}
\bibitem{Cherenkov} P. Cherenkov, Dokl. Akad. Nauk SSSR {\bf2}, 451 (1934).

\bibitem{Tamm} I. Frank and I. Tamm, Dokl. Akad. Nauk SSSR {\bf14}, 107 (1937).

\bibitem{Ypsilantis} T. Ypsilantis and J. S\'eguinot, Nucl. Instrum. Meth. A, {\bf 343}, 30 (1994).

\bibitem{Abashian} A. Abashian, K. Gotow, N. Morgan, L. Piilonen, S. Schrenk, K. Abe, I. Adachi, J. Alexander, K.Aoki and S. Behari, Nucl. Instrum. Meth. A, {\bf 479}, 117 (2002).

\bibitem{Adam} I. Adam, R. Aleksan, L. Amerman, E. Antokhin, D. Aston, P. Bailly, C. Beigbeder, M. Benkebil, P. Besson and G. Bonneaud, Nucl. Instrum. Meth. A, {\bf 538}, 281 (2005).


\bibitem{Adamo} G. Adamo, K. F. MacDonald, Y. Fu, C. Wang, D. Tsai, F. J. Garc\'ia de Abajo and Zheludev, Phys. Rev. Lett. {\bf 103}, 113901 (2009).

\bibitem{Liu} S. Liu, P. Zhang, W. Liu, S. Gong, R. Zhong, Y. Zhang and M. Hu,
 Phys. Rev. Lett. {\bf 109}, 153902 (2012)
 
\bibitem{Ye} F. Liu, L. Xiao, Y. Ye, M. Wang, K. Cui, X. Feng, W. Zhang and Y. Huang, Nat. Photonics {\bf 11}, 289 (2017).


\bibitem{Luo} H. Hu, X. Lin and Y. Luo, Free-electron radiation engineering via structured environments, arXiv: 2108.06149.

\bibitem{Urrutia} R. Mart{\'\i}nez von Dossow, E. Barredo-Alamilla, M. A. Gorlach and L. F. Urrutia, Phys. Rev. D {\bf 113}, 016010 (2026).


\bibitem{Maldacena} "Lectures on AdS/CF" Prepared for Theoretical Advanced
Study Institute in Elementary Particle Physics (TASI $2002$): Particle Physics
and Cosmology: The Quest for Physics Beyond the Standard Model(s), Boulder, Colorado, $2$-$28$ Jun $2002$.

\bibitem{Gabadadze} G. Gabadadze, Phys. Rev. D, {\bf 58} (1998) 094015.

\bibitem{Ansoldi} S. Ansoldi, A. Aurilia, A. Smailagic and E. Spallucci, Phys. Lett. B, {\bf 471} (1999) 133.

\bibitem{Spallucci1} A. Smailagic and E. Spallucci, Phys. Rev. D, {\bf 61}
(2000) 067701.

\bibitem{Spallucci2} A. Smailagic and E. Spallucci, Phys. Lett. B, {\bf 489}
(2000) 435.

\bibitem{Marinatto} S. Ansoldi, A. Aurilia, L. Marinatto and E. Spallucci, Prog. Theor. Phys., {\bf 103} (2000) 1021.

\bibitem{Spallucci3} A. Smailagic and E. Spallucci, J. Phys. A, {\bf 34} (2001)
L435.

\bibitem{Aurilia} A. Aurilia, P. Gaete, J. A. Helay\"el-Neto and E. Spallucci,
EPL, {\bf 117} (2017) 61001. 

\bibitem{PGaete} P. Gaete and E. Spallucci, Phys. Lett. B {\bf 675}, (2009) 145.

\bibitem{PGaete2} J. P. Ferreira, P. Gaete and J. A. Helay\"el-Neto, Eur. Phys. J. Plus, {\bf 140}, (2025) 418.

\bibitem{Gruzinov} A. Gruzinov, A. Hook, and J. Huang, "The universe as a superconductor: How to search for a photon mass", [hep-ph: 2609.26888].
\end{thebibliography}
\end{document}